# DeepVenn – a web application for the creation of area-proportional Venn diagrams using the deep learning framework Tensorflow.js

Tim Hulsen[1,*]

[1]Department of Hospital Services & Informatics, Philips Research, Eindhoven, the Netherlands

*To whom correspondence should be addressed.

**Abstract**

**Motivation:** The Venn diagram is one of the most popular methods to visualize the overlap and differences between data sets. It is especially useful when it is are 'area-proportional'; i.e. the sizes of the circles and the overlaps are proportional to the sizes of the data sets. There are some tools available that can generate area-proportional Venn Diagrams, but most of them are limited to two or three circles, and others are not available as a web application or accept only numbers and not lists of IDs as input. Some existing solutions also have limited accuracy because of outdated algorithms to calculate the optimal placement of the circles. The latest machine learning and deep learning frameworks can offer a solution to this problem.

**Results:** The DeepVenn web application can create area-proportional Venn diagrams of up to ten sets. Because of an algorithm implemented with the deep learning framework Tensorflow.js, DeepVenn automatically finds the optimal solution in which the overlap between the circles corresponds to the sizes of the overlap as much as possible. The only required input is two to ten lists of IDs. Optional parameters include the main title, the subtitle, the set titles and colours of the circles and the background. The user can choose to display absolute numbers or percentages in the final diagram. The image can be saved as a PNG file by right-clicking on it and choosing "Save image as". The right side of the interface also shows the numbers and contents of all intersections.

**Availability:** DeepVenn is available at https://www.deepvenn.com.

**Contact:** tim.hulsen@philips.com

## 1 Introduction

In data science, it is often useful to see the overlap between different data sets, in terms of patient IDs, gene names, etc. One of the most popular methods to visualize the overlap between data sets is the Venn diagram: a diagram consisting of two or more circles in which each circle corresponds to a data set, and the overlap between the circles corresponds to the overlap between these data sets. Venn diagrams are especially useful when they are 'area-proportional' i.e. the sizes of the circles and the overlaps correspond to the sizes of the data sets. In 2007, the BioVenn web interface (Hulsen, et al., 2008) was launched, which is being used by a growing number of researchers (Scholar, 2020). The web interface was followed by R and Python packages using the same algorithm (Hulsen, 2021). However, BioVenn can only generate Venn Diagrams of two or three circles. The number of programs that can generate area-proportional diagrams of more than three circles is limited. The venneuler R package and Cytoscape plugin (Wilkinson, 2012) can do it, providing a 'statistically-justifiable approximation' for the more complicated cases. Unfortunately, venneuler is not available as a web interface, restricting it to users of R and Cytoscape. The nVenn R package (Perez-Silva, et al., 2018) can create 'n-dimensional, quasi-proportional' Venn diagrams, but it has its limitations as well: "However, representing more than six sets takes a long time and is hard to interpret, unless many of the regions are empty" (Quesada, 2020). The tools VennMaster (Kestler, et al., 2008) and DrawEuler (Stapleton, et al., 2011) were apparently able to generate area-proportional Venn diagrams of more than three circles, but they do not seem to be available anymore. The only tool that resembles DeepVenn in terms of functionality is the web application eulerr.co (Larsson, 2020) (and its R package eulerr (Larsson, 2020)), but it only accepts numbers and not lists of IDs as input. It can also be rather slow when a more complicated diagram needs to be created.

## 2   Methods

DeepVenn uses the Tensorflow.js API (TensorFlow, 2020), which is a client-side Javascript implementation of the machine learning platform. It uses a stochastic gradient descent (SGD) algorithm in which the loss is determined by the difference between the calculated and observed distance between each two circles. The learning rate is updated during each epoch to come to an optimal solution. The exact steps are as follows:

1) Remove duplicate IDs (note: DeepVenn is case-sensitive)
2) Calculate the intersects of all possible combinations
3) Remove intersects that are a subset of other intersects
4) Display the sets and intersects on the right side of the interface
5) Calculate the radii of the circles so that the areas of the circles correspond to the size of the datasets they represent
6) Calculate the distances between the centers of the circles, so that the areas of the two-circle overlaps correspond to the size of the datasets they represent (see figure 1 of (Hulsen, et al., 2008))
7) Initialize the circles' centers, starting with the biggest circle and continuing with the circles with the shortest distance to the previous circle. At initialization, the circles are placed in a circular order on the screen.
8) Run an optimization algorithm using the Stochastic Gradient Descent (SGD) optimizer of Tensorflow.js (TensorFlow, 2020), with as tensors the X and Y coordinates of the center of each of the circles. The loss is calculated as the difference between the observed distance and the calculated distance (step 6). The learning rate has an initial value of 10 times the number of sets, and is updated during each epoch as the loss divided by 10 times the number of sets (with a maximum of 100). This method has been tested to give a very high chance of reaching the global minimum instead of a local minimum for the loss function. During each epoch, the whole image is updated: the circles, texts and colors.
9) The optimization stops when:
   - The loss is smaller than 1; or
   - The user clicks on 'Stop Venn Diagram'; or
   - The loss grows 5 times in a row. In this case, DeepVenn will revert to the epoch with the lowest loss.
10) Place the numbers (absolute or percentages) at the 'pole of inaccessibility' (the point within a polygon that is farthest from an edge, (Agafonkin, 2016)), if the user selected this option.

## 3   Results

The DeepVenn web application can generate area-proportional Venn diagrams of two to ten circles from lists of identifiers. Input parameters include up to ten lists of identifiers, which can be copy-pasted into text boxes on the left side of the interface. The name and colour of the circles can be adjusted there as well. On the right side, the title and subtitle of the Venn diagram can be changed, as well as the background colour. The right side of the interface also shows all sets and intersects with size larger than or equal to 1. After clicking on one of these sets or intersects, the text box on the bottom right shows the identifiers within that set or intersection. The Venn diagram can be saved as PNG simply by right-clicking on it and choose "Save as".

The following input recreates figure 7 of (Wilkinson, 2012), but then with numbers included in the figure. A screenshot of the Venn diagram within the DeepVenn interface can be seen in figure 1.

Set A: A, B, C, D, E, F, G, H, I
Set B: E, F, G, J, K, L, M, N, O, P, Q, R, S, T, U
Set C: P, Q, R, V, W, X, Y
Set D: R, S, Y, Z, AA, AB
Set E: AE, AB, AC, AD, AE, AF, AG, AH, AI, AJ
Set F: G, H, I, T, U, AJ, AK, AL, AM

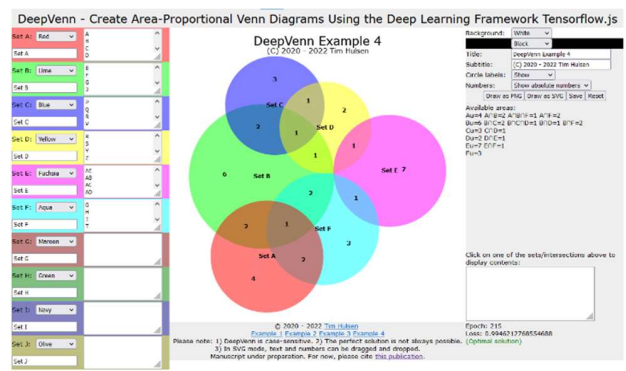

**Fig. 1.  Example diagram.** This example was created by just entering six lists of IDs and setting the 'title' parameter.

The client-side implementation of DeepVenn ensures that possible confidential data (the lists of identifiers) are not stored on the server anywhere. The diagram (which is an HTML5 canvas) can be stored easily as a PNG file by right-clicking on it and choosing 'Save image as'.

## Funding

This work has received no funding.
*Conflict of Interest:* Tim Hulsen is currently employed by Philips Research.